\input{aipcheck}

\documentclass[
    ,final            
  ]
  {aipproc}

\layoutstyle{6x9}

\newcommand{\lsim}{{\, \lower2truept\hbox{
${< \atop\hbox{\raise4truept\hbox{$\sim$}}}$}\,}}
\newcommand{\gsim}{{\, \lower2truept\hbox{
${> \atop\hbox{\raise4truept\hbox{$\sim$}}}$}\,}}

\begin{document}

\title{Supernova and GRB connection: \\
Observations and Questions}

\classification{97.60}
\keywords      {Supernovae, Gamma Rays}

\author{Massimo Della Valle}{
  address={INAF-Arcetri Astrophysical Observatory,  Largo E. Fermi, 5 Firenze, 
Italy\\
Kavli Institute for Theoretical Physics, University of California, Santa Barbara, CA 93106}
}

\begin{abstract}
We review the observational status of the supernova/gamma-ray burst
connection. Present data suggest that SNe associated with GRBs form a
heterogeneous class of objects including both bright and faint
hypernovae and perhaps also `standard' Ib/c events. Evidence for
association with other types of core-collapse SNe (e.g. IIn) is much
weaker. After combining the local GRB rate with the local SN-Ibc rate
and beaming estimates, we find the ratio GRB/SNe-Ibc in the range $\sim
0.5\%-4\%$. In most SN/GRB associations so far discovered, the SN and
GRB events appear to go off simultaneously. In some cases data do not
exclude that the SN explosion may have preceded the GRB by a few
days. Finally we discuss a number of novel questions started by recent
cases of GRB-SN associations.
\end{abstract}

\maketitle 


\section{Introduction}

Intensive optical, infrared and radio follow-up of GRBs, occurring in
the last decade, has established that long-duration GRBs (Klebesadel
1990; Dezalay et al. 1992; Kouveliotou et al. 1993), or at least a
significant fraction of them, are directly connected with supernova
explosions. Most of the evidence arises from observations of supernova
features in the spectra of a few GRB afterglows. Examples of the SN/GRB
connection include SN\,1998bw/GRB\,980425 (Galama et al. 1998),
SN\,2003dh/GRB\,030329 (Stanek et al. 2003, Hjorth et al. 2003),
SN\,2003lw/GRB\,031203 (Malesani et al. 2004), SN\,2002lt/GRB\,021211
(Della Valle et al. 2003), XRF\,020903 (Soderberg et al. 2005),
SN\,2005nc/GRB\,050525A (Della Valle et al. 2006) and more recently
SN\,2006aj/GRB\,060218 (Masetti et al. 2006; Modjaz et al. 2006, Campana
et al. 2006; Sollerman et al. 2006; Pian et al. 2006; Mirabal et
al. 2006; Cobb et al. 2006). In addition there are about a dozen
afterglows which show, days to weeks after the gamma-ray events,
rebrightenings and/or flattenings in their lightcurves (e.g. Zeh et
al. 2004). These bumps are interpreted as SNe emerging out of their
afterglows (Bloom et al. 1999, Castro-Tirado \& Gorosabel 1999). The
detection of star-formation features in the host galaxies of GRBs
(Djorgovski et al. 1998, Fruchter et al. 1999) provided the earliest
hint for the existence of a link between GRBs and the death of massive
stars. Le Floc'h et al. (2003) and Christensen, Hjorth \& Gorosabel
(2004) have found that GRB hosts are galaxies with a fairly high
(relative to the local Universe) star formation of the order of
10~$M_\odot$~yr$^{-1}/L^\star$ or more, while recent studies on the
parent galaxies of GRBs (Conselice et al. 2005, Wainwright et al. 2005)
show that a significant fraction of GRB hosts ($\sim 50\%$) exhibit a
merger/disturbed morphology.

\section{GRB 980425/SN 1998bw}

SN\,1998bw was the first SN discovered spatially and temporally
coincident with a GRB (GRB\,980425; Galama et al. 1998). It was
discovered in the nearby galaxy ESO\,184-G82 at 40 Mpc. This implied
that GRB\,980425 was underenegetic by about 3--4 orders of magnitude
with respect to the ``standard'' $\gamma$-energy budget of $\sim
10^{51}$erg (Frail et al. 2001, Panaitescu \& Kumar 2001). The
associated SN was extremely energetic with expansion velocities 3-4
times higher than those exhibited by normal Ib/c SNe (Patat et
al. 2001). The theoretical modeling of the light curve and spectra
(Iwamoto et al. 1998; Woosley, Eastman, \& Schmidt 1999) suggests that
SN 1998bw originated in a $\sim 40 M_\odot$ star (on the main sequence),
with a C+O core of about $\sim 10 M_\odot$. This picture is supported by
the radio properties of SN 1998bw (Kulkarni et al. 1998, Weiler et
al. 2002) that can be explained in terms of an interaction of a mildly
relativistic ($\Gamma \sim$ 1.8) shock with a dense circumstellar medium
fed by the strong wind of the massive envelope-stripped
progenitor. Recently, Maeda et al. (2006), after analyzing the Fe and
[O\,I] line profiles in the nebular spectra of SN 1998bw, give support
to the idea that SN 1998bw was the product of an asymmetric explosion
(see also H\"oflich, Wheeler \& Wang 1999) viewed relatively off-axis
from the jet direction, $\theta < 30^\circ$ (Maeda et al. 2006) and
$\theta > 15^\circ$ (Yamazaki et al. 2003). An asymmetric explosion
decreases the kinetic energy input by a factor $\lsim 3$ (in Maeda et
al. models) with respect to spherically symmetric models.

However, the association between two peculiar astrophysical objects,
such as SN 1998bw and GRB 980425 was not taken as proof of a general
SN/GRB connection.

\section{GRB~021211/SN 2002lt}

GRB\,021211 was detected by the \mbox{HETE--2} satellite (Crew et
al. 2003), allowing the localization of its optical afterglow (Fox et
al. 2003) and the measurement of its redshift $z=1.006$ (Vreeswijk et
al. 2006). Figure~1 shows the result of the late-time photometric
follow-up, carried out with the ESO VLT--UT4 (Della Valle et al. 2003),
together with observations collected from literature. A rebrightening is
apparent, starting $\sim 15$~days after the burst and reaching the
maximum ($R \sim 24.5$) during the first week of January. For
comparison, the host galaxy has a magnitude $R = 25.22
\pm 0.10$, as measured in late-time images. 
The spectrum of the bump (Fig. 1, right panel) obtained during the
rebrightening phase is characterized by a broad absorption, the
minimum of which is measured at $\sim 3770$~\AA{} (in the rest frame
of the GRB). The comparison with the spectra of other SNe supports the
identification of the broad absorption with a blend of the Ca\,II $H$
and $K$ absorption lines. The blueshifts corresponding to the minimum
of the absorption and to the edge of the blue wing imply velocities $v
\sim 14\,400$~km/s and $v \sim 23\,000$~km/s, respectively. 
In Fig.~1 the light curve of SN\,1994I (dereddened by $A_V = 2$~mag) is
added to the afterglow and host contributions, after applying the
appropriate $K$-correction (solid line). As it can be seen, this model
reproduces well the shape of the observed light curve. It is interesting
to note that SN\,1994I (the spectrum of which provides the best match to
the observations) is a ``standard'' type-Ic event (Filippenko et
al. 1995) rather than a bright {\sl hypernova} (HN) (a hypernova is a
broad-line type-Ibc SN) as the ones proposed for association with other
long-duration GRBs. However we note that even if the pre-maximum spectra
of some HNe (e.g. 2002ap, Mazzali et al. 2002) show significantly
broader lines than our case, this difference vanished after maximum,
such that it may not be easy to distinguish at later stages between the
two types of SNe.

\begin{figure}[!h]
\includegraphics [width=6.2cm,keepaspectratio,angle=0]{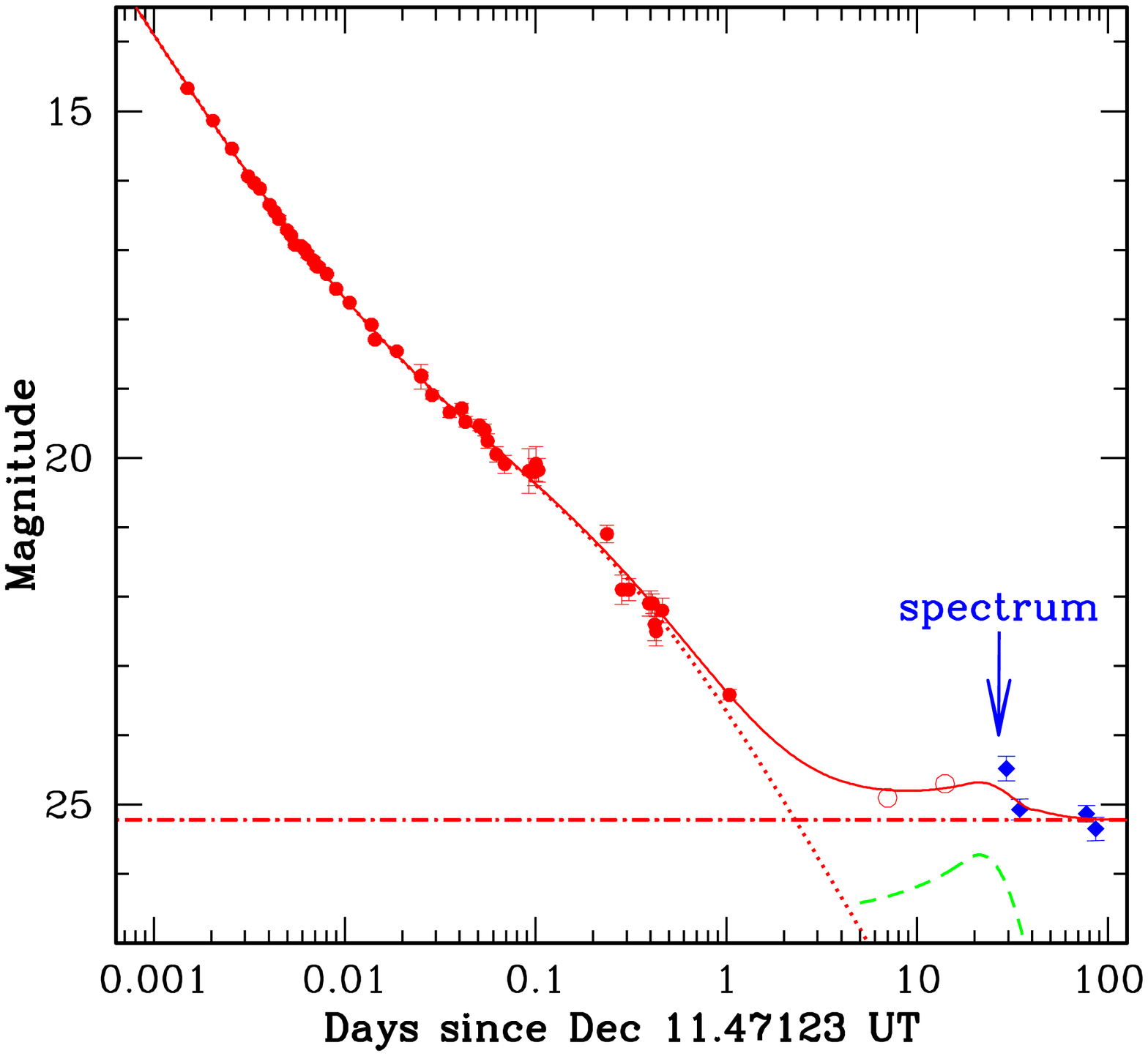}
\includegraphics [width=6.9cm,keepaspectratio,angle=0]{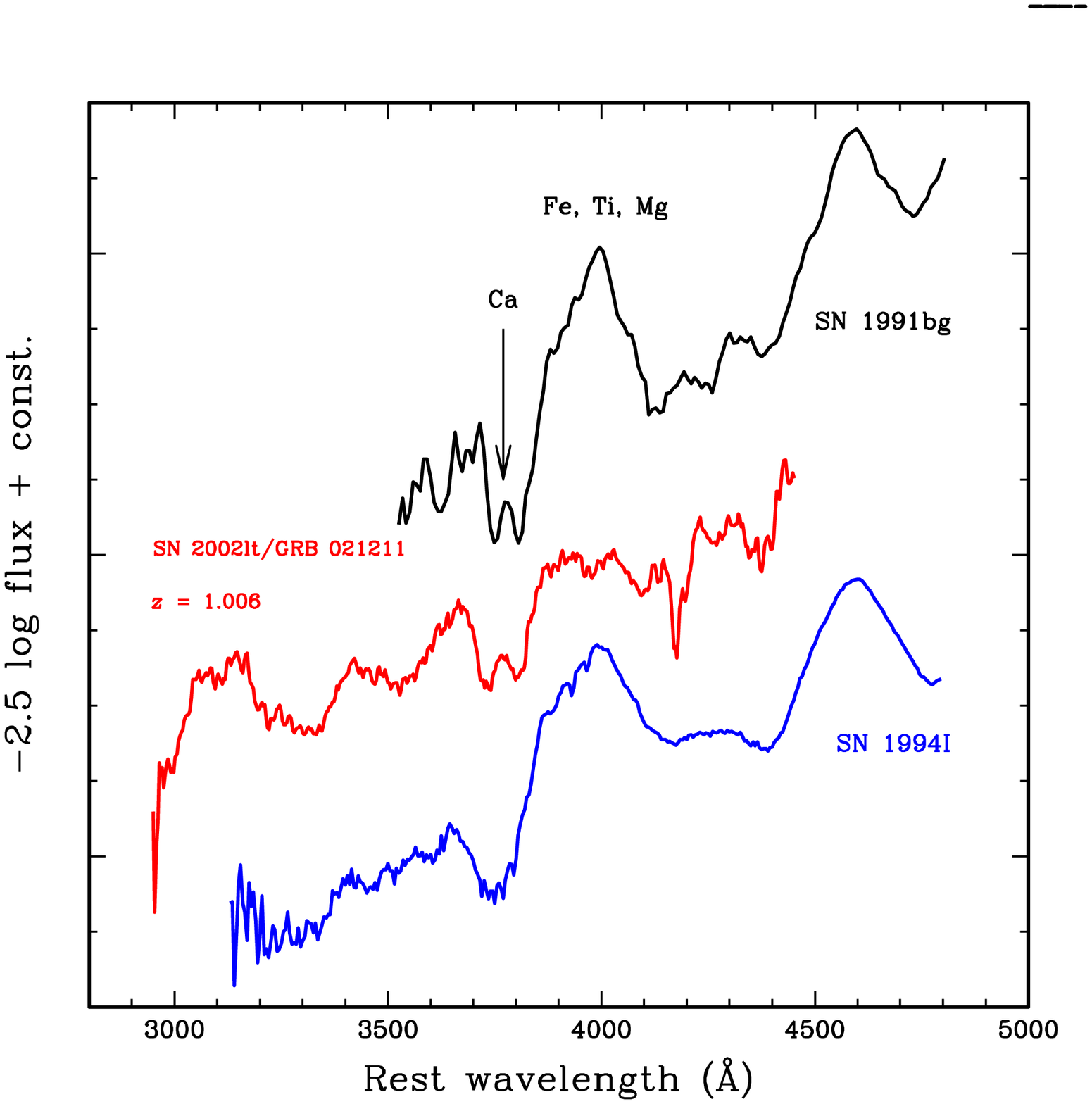}

\caption{{\bf Left panel.} Light curve of the afterglow of GRB 021211. 
Filled circles represent data from published works (Fox et al. 2003; Li
et al. 2003; Pandey et al. 2003), open circles are converted from HST
measurements (Fruchter et al. 2002), while filled diamonds indicate our
data; the arrow shows the epoch of our spectroscopic measurement. The
dotted and dot-dashed lines represent the afterglow and host
contributions respectively. The dashed line shows the light curve of SN
1994I reported at $z=1.006$ and dereddened with $A_V=2$ (from Lee et
al. 1995). The solid line shows the sum of the three contributions.
{\bf Right panel.} Spectrum of the afterglow+host galaxy of GRB 021211
(middle line), taken on 2003 Jan 8.27 UT (27 days after the burst). For
comparison, the spectra of SN 1994I (type Ic, bottom) and SN 1991bg
(peculiar type Ia, top) are displayed, both showing the Ca
absorption. Plots from Della Valle et al. (2003, 2004).}
\label{figura1}
\end{figure}

\section{The ``Smoking Gun'': GRB\,030329/SN\,2003dh}

The breakthrough in the study of the GRB/SN association arrived with
the bright GRB\,030329. This burst, also discovered by the
\mbox{HETE--2} satellite, was found at a redshift $z = 0.1685$
(Greiner et al. 2003). SN features were detected in the spectra of the
afterglow by several groups (Stanek et al. 2003, Hjorth et al. 2003;
see also Kawabata et al. 2003; Matheson et al. 2003a) and the
associated SN (SN\,2003dh) looked strikingly similar to SN\,1998bw
(Fig. 2). The gamma-ray and afterglow properties of this GRB were not
unusual among GRBs, and therefore, the link between GRBs and SNe was
eventually established to be general.

\begin {figure}[!h]
\includegraphics [width=6.9cm, height=8.2cm,angle=90]{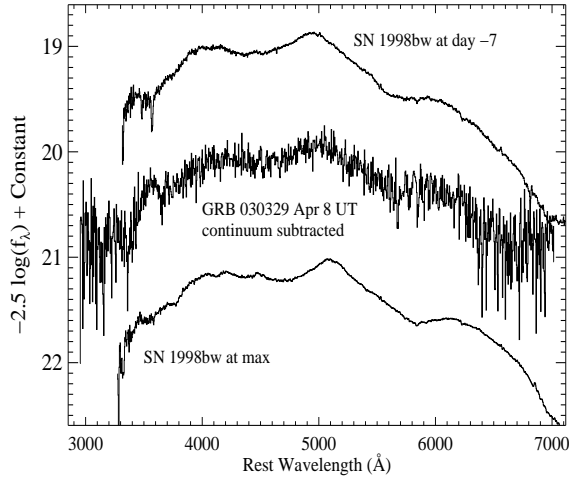}
\caption{Spectrum of SN\,2003dh taken on 2003 April 8, afterg
sbutracting the spectrum of April 4 rescaled. The residual spectrum
shows broad bumps at approximately 5000 and 4200 \AA~ (rest frame),
which is similar to the spectrum of the peculiar type-Ic SN 1998bw a
week before maximum light (Patat et al. 2001). Plot from Stanek et
al. 2003.}
\label{figura2}
\end{figure}

The modeling of the early spectra of SN 2003dh (Mazzali et al. 2003) has
shown that SN 2003dh had a high explosion kinetic energy, $\sim 4\times
10^{52}$ erg (if spherical symmetry is assumed). However, the light
curve derived from fitting the spectra suggests that SN 2003dh was
probably fainter than SN 1998bw (but see Bloom et al. 2004), ejecting
only $\sim 0.35$ M$_\odot$ of $^{56}$Ni. The progenitor was a massive
envelope-stripped star of $\sim 35-40M_\odot$ on the main sequence. The
spectral analysis of the nebular-phase emission lines carried out by
Kosugi et al. (2004) suggests that the explosion of the progenitor of
GRB 030329 was aspherical, and that its axis was well aligned with both
the GRB relativistic jet and our line of sight.

\section{GRB\,031203/SN\,2003lw}

GRB\,031203 was a 30s burst detected by the INTEGRAL burst alert system
(Mereghetti et al. 2003) on 2003 Dec 3. At $z = 0.1055$ (Prochaska et
al. 2004), it was the second closest burst after GRB\,980425. The burst
energy was extremely low, of the order of $10^{49}$~erg, well below
$\sim 10^{51}$~erg of normal GRBs. In this case, a very faint NIR
afterglow was discovered, orders of magnitude dimmer than usual GRB
afterglows (Malesani et al. 2004). A few days after the GRB, a
rebrightening was apparent in all optical bands (Thomsen et al. 2004;
Cobb et al. 2004; Gal-Yam et al. 2004).  For comparison, in Fig. 3 the
$VRI$ light curves of SN\,1998bw are plotted (solid lines), placed at $z
= 0.1055$, stretched by 1.1 and dereddened with $E_{B-V} = 1.1$. After
assuming a light curve shape similar to SN\,1998bw, which had a rise
time of 16~days in the $V$ band, data suggest an explosion time nearly
simultaneous with the GRB. With the assumed reddening, SN\,2003lw
appears to be brighter than SN\,1998bw by 0.5~mag in the $V$, $R$, and
$I$ bands. The absolute magnitudes of SN\,2003lw are hence $M_V =
-19.75\pm0.15$, $M_R = -19.9\pm0.08$, and $M_I = -19.80\pm0.12$. Fig. 3
also shows the spectra of the rebrightening on 2003 Dec ~20 and Dec ~30
(14 and 23 rest-frame days after the GRB), after subtracting the
spectrum taken on Mar ~1 (81 rest-frame days after the GRB, Tagliaferri
et al. 2004). The spectra of SN\,2003lw are remarkably similar to those
of SN\,1998bw obtained at comparable epochs (shown as dotted lines in
Fig.~3). Both SNe show very broad absorption features, indicating high
expansion velocities. The analysis of early spectra of 2003lw (Mazzali
et al. 2006) indicates that this HN produced about $\sim 0.5 M_\odot$ of
Ni. The progenitor mass could be as large as 40-50 $M_\odot$ on the main
sequence.
\bigskip

\begin {figure}[!h]
\includegraphics [ width=5.9cm, height=7.2cm, angle=0]{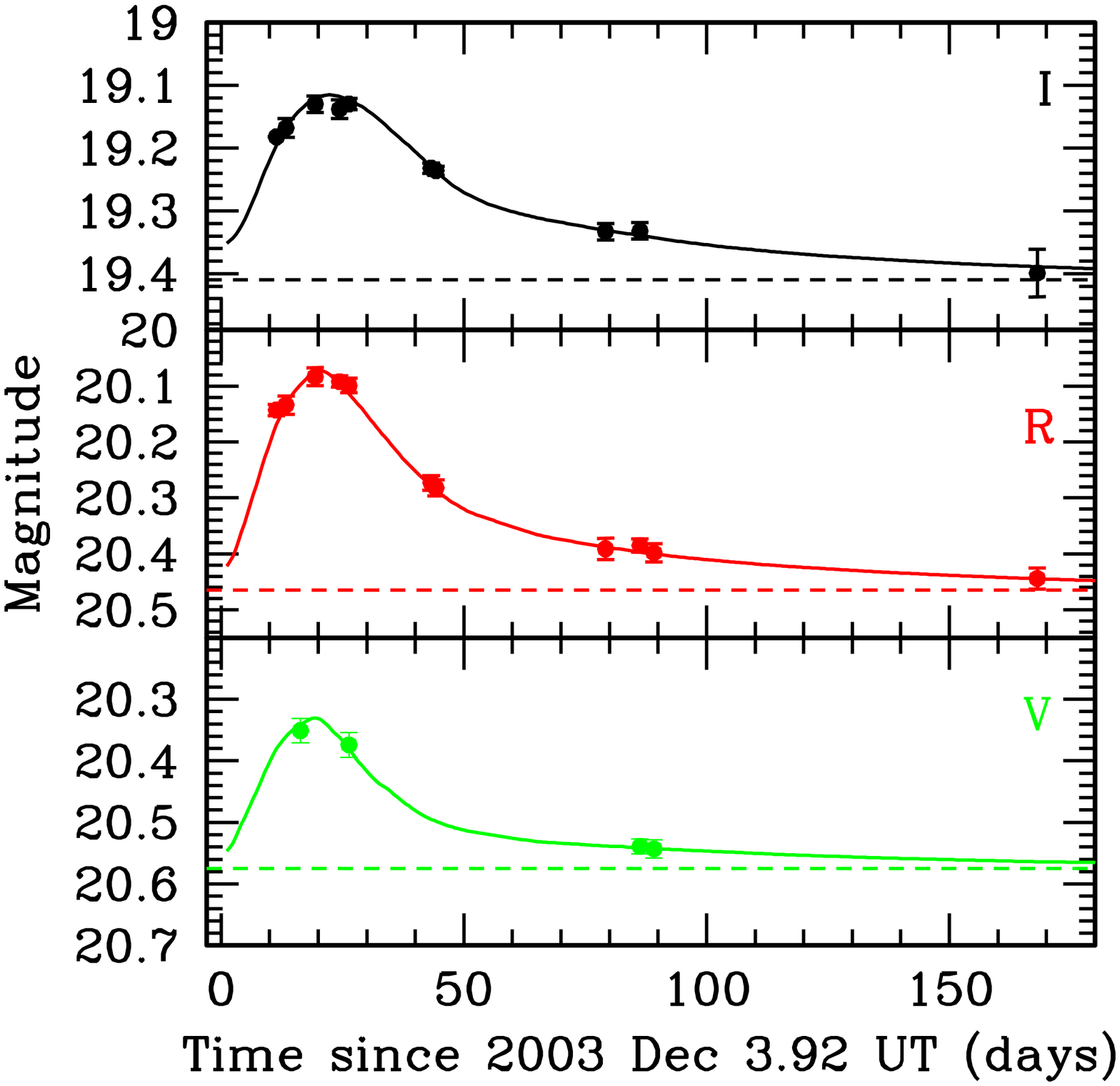}
\includegraphics [width=6.9cm, height=8.2cm,angle=0]{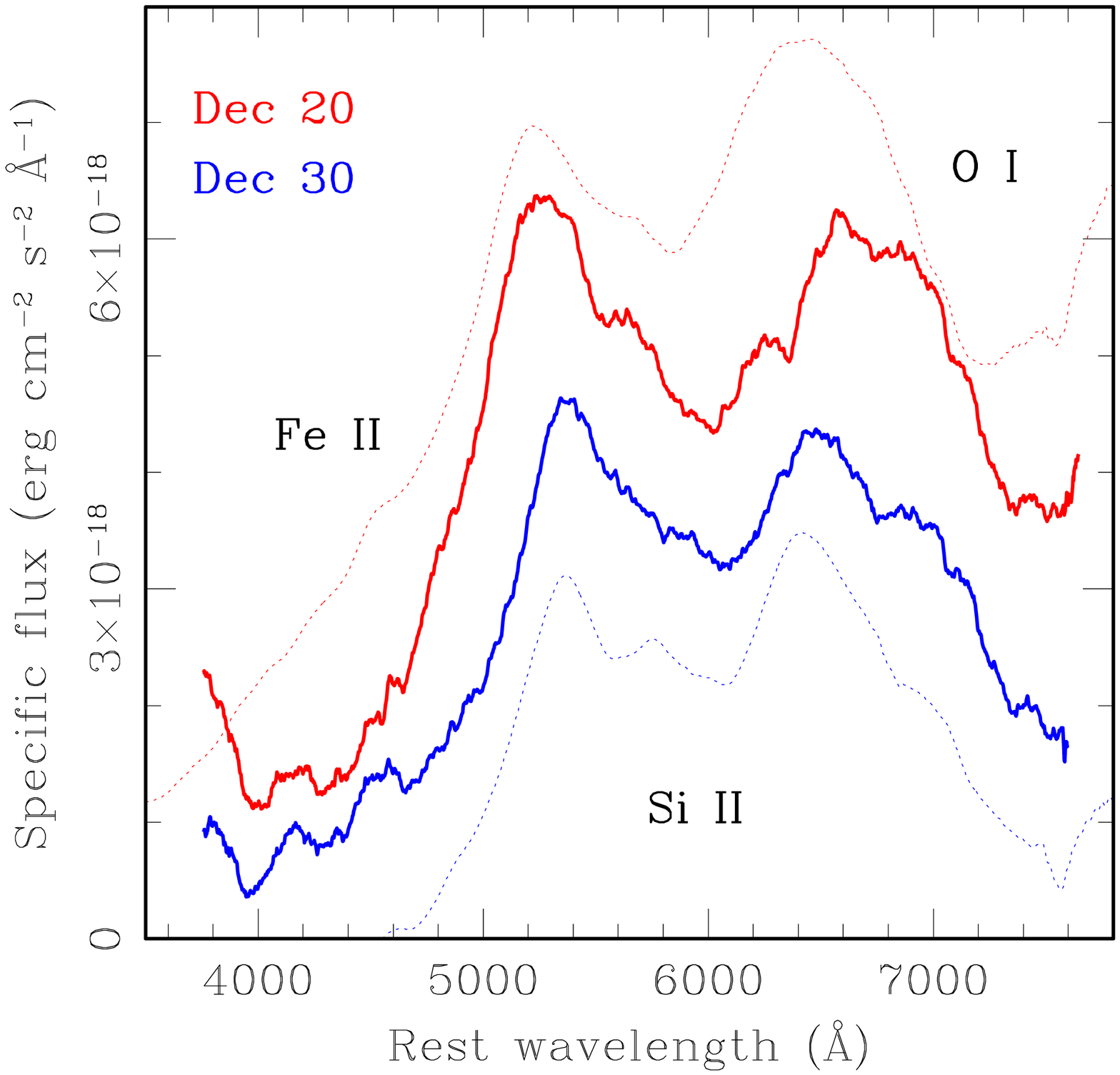}
\caption{{\bf Left panel.} 
Optical and NIR light curves of GRB 031203 (dots). The solid curves show
the evolution of SN 1998bw (Galama et al. 1998; McKenzie \& Schaefer
1999), rescaled at $z=0.1055$, stretched by a factor 1.1, extinguished
with $E(B-V)=1.1$, and brightened by 0.5 mag. The dashed lines indicate
the host galaxy contribution. {\bf Right panel.}  Spectra of SN 2003lw,
taken on 2003 December 20 and 30 (solid lines), smoothed with a boxcar
filter 250\AA{} wide. Dotted lines show the spectra of SN 1998bw (from
Patat et al. 2001), taken on 1998 May 9 and 19 (13.5 and 23.5 days after
the GRB, or 2 days before and 7 days after the $V$-band maximum,
respectively), extinguished with $E(B-V)=1.1$ and a Galactic extinction
law (Cardelli et al. 1989). The spectra of SN 1998bw were vertically
displaced for presentation purposes. Plots from Malesani et al. 2004.}
\label{figura3}
\end{figure}

\begin{table}
\begin{tabular}{crr}
\hline
  \tablehead{1}{r}{b}{SN}
& \tablehead{1}{r}{b}{cz~km/s}
& \tablehead{1}{r}{b}{References}\\
\hline
     1997dq & 958 &  Mazzali et al. 2004\\
     1997ef & 3539&  Filippenko 1997b\\
     1998bw & 2550&  Galama et al. 1998 \\
     1999as & 36000& Hatano et al. 2001\\		
     2002ap & 632 &  Mazzali et al. 2002, Foley et al. 2003\\ 
     2002bl & 4757&  Filippenko et al. 2002\\
     2003bg & 1320&  Filippenko \& Chornack 2003\\
     2003dh & 46000& Stanek et al. 2003, Hjorth et al. 2003\\	
     2003jd & 5635&  Filippenko et al.2003; Matheson et al. 2003b\\
     2003lw & 30000& Malesani et al. 2004\\
     2004bu & 5549&  Foley et al. 2004 \\
     2005kz & 8117&  Filippenko, Foley \& Matheson 2005\\		
     2006aj & 9000&  Masetti et al. 2006\\
    \hline
  \end{tabular}
  \caption{Hypernovae}
\end{table}

\section{Rates of SNe Ib/c, Hypernovae and GRBs}

GRB\,980425, XRF\,020903, GRB\,031203, and GRB\,060218 have
$\gamma$-energy budgets between 2--4 orders of magnitude fainter than
those exhibited by ``standard'' GRBs. The increasing number of discovery
of these underenergetic events (with an associated SN component) can no
longer considered a simple collections of peculiar, atypical cases.
These bursts were so faint, that they would have been easily missed at
cosmological distances, therefore it is very possible that they are the
most frequent GRBs in the universe.  These 4 events have been detected
in 9 years of observations within a volume of $\sim 4$ Gpc$^3$
(XRF\,020903 is the most far away under-energetic event so far observed,
at $z = 0.25$). These figures may imply an ``observed'' rate of $\sim
0.1$ GRB Gpc$^{-3}$ yr$^{-1}$.  On the other hand the ``observed'' rate
has to be corrected to account for the effective full-sky coverage of
the satellites. This task is not so simple to be carried out because the
correction should include different factors (effective field of view,
downtimes\ldots) which are not easy to be quantify. As a conservative
estimate (based on the published technical reports) one can crudely
assume a correction factor of $\gsim 10$. Then the ``observable'' local
rate turns out to be: $\sim 2$ GRB Gpc$^{-3}$ yr$^{-1}$, which is
slightly larger than derived by Schmidt (2001) ($\sim 0.5$ GRB
Gpc$^{-3}$ yr$^{-1}$) and consistent with 1.1 GRB Gpc$^{-3}$ yr$^{-1}$)
derived by Guetta et al. (2004).

A rate of $\sim 2.6 \times 10^4$ SNe-Ibc Gpc$^{-3}$ yr$^{-1}$ can be
derived by combining the the local density of $B$-band luminosity of
$\sim 1.2\times 10^8 L_{B,\odot}$ Mpc$^{-3}$ (e.g. Madau, Della Valle \&
Panagia 1998) with the local rate of 0.22 SNe-Ibc (in Irr and Sm Hubble
types) per century and per 10$^{10}$ $L_{B,\odot}$ (Cappellaro, Evans \&
Turatto 1999). This SN rate has to be compared with $\lsim 2$ GRB
Gpc$^{-3}$ yr$^{-1}$ after rescaling for the jet beaming
factor\footnote{The beaming factor is defined as
$f_b^{-1}=1-\cos\theta$}. There exist different estimates for this
parameter: from $\sim 500$ (Frail et al. 2001) to $\sim 75$ (Guetta,
Piran \& Waxman 2005), corresponding to beaming angles $\sim
4^\circ$--$10^\circ$, respectively. Taking these figures at their face
value, we find the ratio GRB/SNe-Ibc to be in the range: $\sim
4\%-0.5\%$. Izzard et al. (2004) have modeled the stellar progenitors of
type-Ibc SNe, selecting those capable to produce GRBs. They find ratios
comparable with the numbers presented here. Radio and optical surveys
give less severe constraints: Soderberg et al. (2006b) and Rau et
al. (2006) find $f_b^{-1} \lsim 10^4$ and $f_b^{-1} \lsim 12500$, then
implying $\theta \gsim 0.8^\circ$ and GRB/SNe-Ibc $\lsim 30\%$.

The computation of the ratio GRB/HN requires a further step. The
measurement of the SN rate is based on the control-time methodology
(Zwicky 1938) that implies the systematic monitoring of galaxies of
known distances. Unfortunately all HNe so far discovered (see Tab. 1)
have been not found during ``time controlled'' surveys. An heuristic
approach to derive the rate of HNe is to compute the frequency of
occurrence of all SNe-Ib/c and HNe in a limited distance sample of
objects and to assume that they have been efficiently (or inefficiently)
monitored by the same extent. From an upgraded version of Asiago catalog
we have extracted 91 SNe-Ib/c (8 of which are HNe) with $cz< 6000$
km/s. This velocity threshold is suitable to make the distance
distribution of `normal' Ib/c and HNe statistically indistinguishable
(KS probability=0.42). After excluding SN 1998bw, because it was
searched in the error-box of GRB 980425, one can infer that the fraction
of HNe is about $7/91\simeq 8\%$ of the total number of SNe
Ib/c. Therefore the ratio GRB/HNe turns out to be $\sim 0.5\div 0.06$
(cfr. $\sim 1$, Podsiadlowski et al. 2004).

Finally we like to stress two points: {\sl a)} sub-energetic GRBs may
be less collimated than classical events. Their low energy demand does
not pose any problem for most progenitor models, thus the beaming
factor does not need to be too large. Moreover, from the sparse data
on their afterglows, it appears that their breaks are quite late, if
existent. For example, interpreting the break in the X-ray light curve
of GRB 031203 as due to a jet we obtain $\theta \sim 16^{\circ} \div
30^{\circ}$. Therefore it is likely that small GRB/SN ratios are
favored; {\sl b)} the estimate of the ``local'' rate of GRBs is
seriously plagued by the small number of available events. For
example, if we consider only the 3 nearest events (which have occurred
within 0.4 Gpc$^3$) we cannot exclude values as high as $\sim 15$ GRB
Gpc$^{-3}$ yr$^{-1}$. In this case the ratio GRB/SNe-Ibc becomes $\sim
30\%-4\%$ (for $f_b=500$ and 75, respectively). The former value would
imply that a significant fraction of ``standard'' SNe-Ibc contributes
to originate GRBs. The latter value might be consistent with a ratio
GRB/HNe $\sim 1$.

\begin{table}
\begin{tabular}{rrrcr} 

  \tablehead{1}{r}{b}{GRB}
& \tablehead{1}{r}{b}{SN}
& \tablehead{1}{r}{b}{$+ \Delta {\rm t(days)}$ }
& \tablehead{1}{r}{b}{$- \Delta {\rm t(days)}$ }
& \tablehead{1}{r}{b}{references} \\
\hline 
GRB 980425 & 1998bw & 0.7 &  2 & Iwamoto et al. 1998\\ 
GRB 000911 & bump & 1.5 & 7 & Lazzati et al. 2001\\ 
GRB 011121 & 2001ke & 0 & 5 & Bloom et al. 2002b \\
           &      & --  & a few & Garnavich et al. 2003\\  
GRB 021211 & 2002lt & 1.5 & 3 & Della Valle et al. 2003\\ 
GRB 030329 & 2003dh & 2 &  8  & Kawabata et al. 2003\\ 
           &        & --  &  2  & Matheson et al. 2003a\\ 
GRB 031203 & 2003lw & 0 & 2 & Malesani et  al. 2004\\ 
GRB 041006 & bump   & 2.7& 0.9& Stanek et al. 2005\\ 
GRB 050525A& 2005nc & 0 & $<3.5$ & Della Valle et al. 2006\\
GRB 060218 & 2006aj & 0 & 0    & Campana et al. 2006\\
\hline 
\end{tabular}

\caption{Supernova/gamma-ray burst time lag. A negative time lag indicates
  that the SN explosion precedes the GRB.}  
\end{table}

\section{Facts and Open Questions}

From the data presented in the previous sections, a number of both
established facts and intriguing questions emerge:

{\sl 1.} Long duration GRBs are closely connected with the death of
massive stars. This has been spectroscopically confirmed over a large
range of redshifts: GRB\,980425/SN\,1998bw at $z=0.0085$ (Galama et
al. 1998); GRB\,060218/SN\,2006aj at $z=0.03$ (Campana et al. 2006);
GRB\,031203/SN\,2003lw at $z=0.1055$ (Malesani et al. 2003);
GRB\,030329/SN\,2003dh at $z=0.16$ (Stanek et al. 2003, Hjorth et
al. 2003); XRF\,020903/SN\,1998bw-like at $z=0.23$ (Soderberg et
al. 2005); GRB\,050525A/SN\,2005cn at $z=0.6$ (Della Valle et al. 2006);
and GRB\,021211/SN\,2002lt at $z \sim 1$ (Della Valle et al. 2003). In
spite of this tight connection with SN explosions, Fruchter et al. (2006)
have demonstrated that GRBs and SNe do not occur in similar galactic
environments, these authors argue that this unexpected behavior may be
related to the low metallicity content observed in the GRB hosts.

{\sl 2.} It is not clear whether or not only HNe (i.e. broad-lines
SNe-Ibc) are capable of producing GRBs or even ``standard'' Ib/c
events. There is weak evidence that other type of core-collapse SNe,
such as type IIn, can contribute to the SN population of GRBs (Germany
et al. 2000; Turatto et al. 2000; Rigon et al. 2003). However, in a
recent study, Valenti et al. (2005) (see also Bosnjak et al. 2006) were
not able to corroborate, on statistical basis, the associations with
core-collapse SNe different from SNe-Ibc. The best evidence for the case
of an association between a type-IIn SN and a GRB has been provided by
Garnavich et al. (2003), who found that the color evolution of the bump
associated with GRB\,011121 is consistent with the color evolution of an
underlying SN (SN\,2001ke) strongly interacting with a dense
circumstellar gas due to the progenitor wind.

{\sl 3.} GRB-SN data (including the bumps: Della Valle et al. 2003,
Fynbo et al. 2004, Levan et al. 2005, Masetti et al. 2003, Price et
al. 2003, Soderberg et al. 2005, Gorosabel et al. 2005, Stanek et
al. 2005, Soderberg et al. 2006a, Bersier et al. 2006) indicate that the
magnitude at maximum of SNe associated with GRBs may span a range of
about 5 magnitudes, which is similar to that exhibited by ``standard''
stripped-envelope stars (Richardson, Branch \& Baron 2006). However all
GRB-SNe which have so far been spectroscopically confirmed appear to
belong to the bright tail of SNe-Ib/c population (all have M$_V \sim
-18.5/-19$). If this is the effect of an observational bias (which
favors the spectroscopic observations of bright SNe) operating on a
small number of objects or it has a deeper physical meaning is not yet
clear.

{\sl 4.} There are events, such as XRF 040701 (Soderberg et al. 2005),
for which the SN has been unsuccessfully searched with HST, down to
magnitude M$_V\sim -15.8 \div -13$ (according to different assumption on
the host galaxy extinction). These observations may imply that some GRBs
can be associated with very underluminous SNe-Ibc. On the other hand
such very faint objects have never been observed (see Richardson, Branch
\& Baron 2006). However, it should be noticed that a few unusually faint
core-collapse events (belonging to the type-II class, not the Ibc!) have
been already observed at magnitudes $M_V \sim -13 \div -14.5$ (Turatto
et al. 1998; Pastorello et al. 2004).  It would be possible that SNe-Ibc
of comparable low luminosity do exist but they have not been observed
just because they are rarer objects than type II (SNe-Ibc are about
15-30\% of type II in late spirals/Irr, e.g. Mannucci et al. 2005).

{\sl 5.} Several authors have reported the detection of Fe and other
metal lines in GRB X-ray afterglows (e.g. Piro et al. 1999). If valid
(see Sako et al. 2005 for a critical view) these observations would
have broad implications for both GRB emission models and would
strongly link GRBs with SN explosions. For example, Butler et
al. (2003) have reported the detection in a Chandra spectrum of
emission lines whose intensity and blueshift would imply that a
supernova occurred $>2$ months prior to the $\gamma$ event. This kind
of observations can be accommodated in the framework of the {\sl
supranova} model (Vietri \& Stella 1998), where a SN is predicted to
explode months or years before the $\gamma$ burst. In Tab. 2 we have
reported the estimates of the lags between the SN explosions and the
associated GRBs, as measured by the authors of the papers.  After
taking these data at their face value, one can conclude that most SN
and GRB events occur simultaneously, and only in some case the SN may
have preceded the GRB by a few days (at the most). However we note
that Swift has not detected X-ray lines in any afterglows so far
observed.

{\sl 6.} Only a very small fraction of all massive stars are capable of
producing GRBs. SNe-Ibc appear to be the natural candidates because they
have already lost the Hydrogen envelope when the collapse of the core
occurs, then allowing the ultra-relativistic jets to escape from the
progenitor star. Nevertheless this fact does not seem to be sufficient.
According to the current SN and GRB rates and $\langle f_b \rangle$
estimates, only $\lsim 4\%$ of type-Ibc SNe are able to produce
GRBs. This implies that GRB progenitors must have some other special
characteristic other than being just massive stars. Recent studies have
extensively discussed the role that stellar rotation (Woosley \& Heger
2006; Yoon \& Langer 2005; Fryer \& Heger 2005), binarity (Podsiadlowski
et al. 2004; Mirabel 2004; Tutukov \& Cherepashchuk 2003; Smartt et
al. 2002), asymmetry (Maeda et al. 2005) and metallicity (e.g. Fruchter
et al. 2006) play in the GRB phenomenon.

{\sl 7.} The ``optical'' properties (i.e. luminosity at peak and
expansion velocities) of the 4 closest SNe associated with GRBs vary by
at most $\sim \pm 50\%$, while the $\gamma$-budget covers about 4 order
of magnitudes. These facts may be interpreted in at least 2 different
ways: a) we may have observed intrinsically similar phenomena under
different angles. GRB 030329/SN 2003dh may be viewed almost pole-on, GRB
980425/SN 1998bw relatively off-axis ($15^\circ<\theta<30^\circ$), while
GRB 031203/SN 2003lw may lie in between (Ramirez-Ruiz et al. 2005). A
consequence of this scenario is that the $\gamma$-properties are
strongly dependent upon the angle ($\sim \theta^4$), whereas the optical
properties are affected much less by changing the viewing angle up to
$\Delta \theta \sim 30^\circ$; b) the recent event
GRB\,060218/SN\,2006aj ($E_{\rm iso} \sim 6\times 10^{49}$ erg), may
suggest a different interpretation. This GRB may be an example of
intrinsically fainter event (Campana et al. 2006; Amati et
al. 2006). This might indicate that there exists an intrinsic dispersion
in the properties of the relativistic ejecta for SNe having similar
optical properties (e.g. peak of luminosity, velocity of the
ejecta). This fact is not unconceivable after keeping in mind that the
observed relativistic energies at play in the GRB phenomenon, at least
in the local universe ($z<0.1$), appear to be just tiny fluctuations
(10$^{-2/-4}$) of the kinetic energy involved in the `standard' SNe-Ibc
($\sim 10^{51}$ erg) or HN explosions ($\sim 10^{52}$ erg).

{\sl 8.} As for AGN, it has been proposed (Lamb et al. 2005, see also
Kouveliotou et al. 2004 and Dado et al. 2004) a unification scheme where
GRBs, XRRs, XRFs and SNe-Ibc are the same phenomenon, but viewed at
different angles. Given the rates of GRBs and type Ibc SNe discussed in
the section 6, the unification scenario would work for $\langle f_b^{-1}
\rangle \sim 30000$, which would correspond to beaming angles of $\sim
0.5^\circ$. On the other hand the measured $f_b^{-1}$ factors are much
smaller, likely in the range 75--500 (Guetta et al. 2005, Yonetoku et
al. 2005; van Putten \& Regimbau 2003, Frail et al. 2001) that
corresponds to beaming angles of $\sim 10^\circ-4^\circ$.
\bigskip
 
{\bf Acknowledgments}
\medskip

I wish to thanks Daniele Malesani, Maurice van Putten and Evan
Scannapieco for the critical reading of the manuscript and all
colleagues of the ``Supernova-Gamma Ray Burst Connection'' program at
the KITP (UCSB) for useful discussions.\\ This research was partially
supported by the National Science Foundation under Grant
No. PHY99-0794.



\bibliographystyle{aipproc}   





\end{document}